\documentclass[cmbright]{mmaauth}

\usepackage{moreverb}

\usepackage[colorlinks,bookmarksopen,bookmarksnumbered,
             citecolor=red,urlcolor=red]{hyperref}

\usepackage{graphicx}        
\DeclareGraphicsExtensions{.pdf}
\usepackage{multicol}        
\usepackage[bottom]{footmisc}
\usepackage{bm}
\usepackage{amsmath}
\usepackage{amssymb}
\usepackage{caption}
\usepackage[noadjust]{cite}

 \newtheorem{thm}{Theorem}

 \newtheorem{example}[thm]{Example}
 
 \numberwithin{equation}{section}

\newcommand{\hf}{\frac{1}{2}}

\newcommand{\gvec}[1]{\ensuremath{\mbox{\textbf{\textit{#1}}}}} 
\newcommand{\sgvec}[1]{\ensuremath{\mbox{\textbf{\textit{\small #1}}}}}
\newcommand{\eo}{e_0}
\newcommand{\einf}{e_{\infty}}
\newcommand{\be}{\begin{equation}}
\newcommand{\ee}{\end{equation}}

  \newcommand{\R}{\ensuremath{\mathbb{R}}}
  \newcommand{\C}{\ensuremath{\mathbb{C}}}

  \newcommand{\Z}{\ensuremath{\mathbb{Z}}}
   
\newcommand\BibTeX{{\rmfamily B\kern-.05em \textsc{i\kern-.025em b}\kern-.08em
T\kern-.1667em\lower.7ex\hbox{E}\kern-.125emX}}

\def\volumeyear{2010}

\begin{document}

\runninghead{E.~Hitzer}

\title{Crystal planes and reciprocal space in Clifford geometric algebra}

\author{E.~Hitzer\corrauth}

\address{Department of Applied Physics, University of Fukui, 3-9-1 Bunkyo, 910-8507 Fukui, Japan}

\corraddr{Department of Applied Physics, University of Fukui, 3-9-1 Bunkyo, 910-8507 Fukui, Japan}

\begin{abstract}
This paper discusses the geometry of $k$D crystal cells given by $(k+1)$ points in a projective space $\R^{n+1}$. We show how the concepts of barycentric and fractional (crystallographic) coordinates, reciprocal vectors and dual representation are related (and geometrically interpreted) in the projective geometric algebra $Cl(\R^{n+1})$ (see \cite{HG:AL1844}) and in the conformal algebra $Cl(\R^{n+1,1})$. The crystallographic notions of $d$-spacing, phase angle, structure factors, conditions for Bragg reflections, and the interfacial angles of crystal planes are obtained in the same context.
\end{abstract}

\MOS{74E15; 15A66}


\maketitle

\vspace{-6pt}

\section{Introduction}

Geometric algebra \cite{DFM:GAfCS} has been successfully applied to the description of crystallographic symmetry \cite{HH:SGinGA,HP:ICCA8}. We now extend this treatment by studying the description of offset subspaces in the geometric algebra of projective space $Cl(\R^{n+1})$ and in the conformal model of Euclidean geometry, i.e. in $Cl(\R^{n+1,1})$, see e.g. \cite{LH:IAaGR}. In particular crystal planes in any dimension are such offset subspaces. Reciprocal vectors appear as support vectors of crystal planes, identical to the Euclidean parts of dual vectors describing hyperplanes in the above model algebras. Regarding crystallography, we assume e.g. familiarity with a modern textbook like \cite{MMJ:FCCA}. 

We mainly address crystallographers, who want to know how to successfully express their problems in the new comprehensive mathematical language of Clifford geometric algebra. We see a need for this, because so far many main stream crystallographers are relatively unfamiliar with Clifford geometric algebra \cite{EH:MaThCrystECM26}. We focus on expressing key notions in standard crystallography, that clearly demonstrate how to employ the powerful invariant and dimension independent tools of Clifford geometric algebra. So far there is no literature on e.g. how to turn crystallographic \textit{Miller indexes} into multivector expressions for crystallographic planes, and based on this how to directly compute crystal plane $d$-spacing, phase angles of structure factors, reflection conditions for the occurrence of Bragg reflections, interfacial angles, and the like. 

In this paper the first section introduces Clifford geometric algebra, the geometric algebra of projective space, and the so called conformal model. Then follow sections on crystal planes in the geometric algebra of projective space, generalizations to higher dimensions, crystal hyperplanes (including Miller indexes and $d$-spacing), and finally crystal planes in the conformal model of Euclidean geometry. This last section gives formulas for how to compute phase angles of structure factors, integral, zonal and serial conditions for Bragg reflections, and interfacial angles.

\section{Clifford geometric algebra, geometric algebra of projective space and conformal model}
\subsection{Clifford geometric algebra}

We first define the notion of a Clifford geometric algebra in plain mathematical terms. \cite{FM:ICNAAM2007}
Let $\{e_1, e_2, \ldots , e_q, e_{q+1}, \ldots , e_n \}$, with $n=p+q$, $e_k^2=\varepsilon_k$, $\varepsilon_k = +1$ for $k=1, \ldots , q$, $\varepsilon_k = -1$ for $k=q+1, \ldots , n$,  be an orthonormal base of the normed vector space $\R^{p,q}$ with a product according to the multiplication rules 
\be
  e_k e_l + e_l e_k = 2 \varepsilon_k \delta_{k,l}, \,\,\,k,l = 1, \ldots n,
\label{eq:mrules}
\ee
where $\delta_{k,l}$ is the Kronecker symbol with $\delta_{k,l}= 1$ for $k=l$, and $\delta_{k,l}= 0$ for $k\neq l$. This non-commutative product generates the associative $2^n$-dimensional Clifford geometric algebra $Cl(\R^{p,q})=Cl(p,q)=Cl_{p,q} = \mathcal{G}_{p,q}$ over $\R$. The set $\{ e_A: A\subseteq \{1, \ldots ,n\}\}$ with $e_A = e_{h_1}e_{h_2}\ldots e_{h_r}$, $1 \leq h_1< \ldots < h_r \leq n$, $e_{\emptyset}=1$, forms a graded basis of $Cl_{p,q}$. The grades $r$ range from $0$ for scalars, $1$ for vectors, $2$ for bivectors, $k$ for $k$-vectors, up to $n$ for so called pseudoscalars. 
The real vector space $\R^{p,q}$ will be embedded in $Cl_{p,q}$ by identifying the element $(a_1,a_2,\ldots,a_n)\in\R^n$ with the element $a=a_1e_1+a_2e_2+\ldots a_ne_n$ of the algebra. The general elements of $Cl_{p,q}$ are real linear combinations of basis blades $e_A$, and are called Clifford numbers, multivectors or hypercomplex numbers. 

The parts of grade $0$, $(s-r)$, $(r-s)$, and $(s+r)$, respectively, of the geometric product of an $r$-vector $A_r\in Cl_{p,q}$ with an $s$-vector $B_s\in Cl_{p,q}$ 
\be 
\langle A_r B_s \rangle_{0} = A_r \ast B_s, \quad
\langle A_r B_s \rangle_{s-r} = A_r \rfloor B_s, \quad
\langle A_r B_s \rangle_{r-s} = A_r \lfloor B_s, \quad
\langle A_r B_s \rangle_{r+s} = A_r \wedge B_s,
\label{eq:gaprods}
\ee 
are called scalar product, left contraction, right contraction, and (associative) outer product, respectively. These definitions extend by linearity to the corresponding products of general multivectors. The various derived products of \eqref{eq:gaprods} are related to each other, e.g. by
\be 
  (A\wedge B)\rfloor C = A\rfloor (B\rfloor C),
  \qquad
  \forall A,B,C \in Cl_{p,q}.
  \label{eq:ABCrel}
\ee 
 Note that for vectors $a,b$ in $\R^{p,q} \subset Cl_{p,q}$ we have
\be 
ab = a \rfloor b + a \wedge b, \quad
a \rfloor b = a \lfloor b = a \cdot b = a \ast b,
\ee 
where $a \cdot b$ is the inner product of $\R^{p,q}$. The geometric interpretation of the bivector $a \wedge b=-b \wedge a$ is an oriented parallelogram area in space with sense ($\pm$ sign), compare Fig. \ref{fg:fig3}. Higher order outer products (blades) $A_r=a_1\wedge \ldots \wedge a_r$ of $r$ linearly independent vectors $a_1, \ldots, a_r \in \R^{p,q}$, $1\leq r \leq n$, are interpreted as oriented $r$-dimensional parallelepipeds in space with orientation and sense. For non zero $A_r^2 = A_r \ast A_r \in \R\setminus \{0\}$, we can define the (right and left) inverse blade $A_r^{-1} = A_r/(A_r^2)$. For example every non-isotropic vector $b \in \R^{p,q}$, $\varepsilon_b|b|^2 = b^2 \neq 0$, $\varepsilon_b = \mathrm{sign}(b^2)$ has inverse $b^{-1} = b/(b^2)$.

The projection and rejection of vector $a$ onto (from) the non-isotropic vector $b$, are defined as
\be 
  P_{b}(a) = (a\rfloor \frac{b}{|b|})\frac{b}{\varepsilon_b|b|} = (a\rfloor b)b^{-1}, 
  \qquad
  P_{b}^{\perp}(a) = a - P_{b}(a) = (ab - a\rfloor b)b^{-1} = (a\wedge b)b^{-1},
\ee 
respectively.
This can be generalized to projections and rejections of blades $A \in Cl_{p,q}$ onto (from) non-isotropic blades $B \in Cl_{p,q}$
\be 
  P_{B}(A) = (A\rfloor B)B^{-1}, 
  \qquad
  P_{B}^{\perp}(A) = (A\wedge B)B^{-1},
\ee
respectively.

All vectors $b$ parallel to a non zero vector $a\in \R^{p,q}$ span a zero parallelogram area with $a$, i.e. the line space spanned by $a\in \R^{p,q}$ is given by $V(a) = \{b\in \R^{p,q}: b\wedge a = 0 \}$. Similarly a subspace of $\R^{p,q}$ spanned by $r$, $1\leq r \leq n$, linearly independent vectors $a_1, \ldots, a_r \in \R^{p,q}$,  is given by $V(a_1, \ldots, a_r) = \{b\in \R^{p,q}: b\wedge a_1\wedge \ldots \wedge a_r = 0 \}$. This subspace representation is called outer product null space representation (OPNS).

The \textit{duality} operation is defined as multiplication by the unit inverse pseudoscalar $I^{-1}=I/(I^2)$ (of maximum grade $n$) of the geometric algebra $Cl_{p,q}$. 
Given an $r$-dimensional subspace $V(a_1, \ldots, a_r)\in \R^{p,q}$ specified by its OPNS representation blade $A_r=a_1\wedge \ldots \wedge a_r$, then its \textit{dual} representation (as inner product null space [IPNS]) is given by the $(n-r)$-blade 
\be 
  A_r^{\ast} = A_r{I}^{-1} 
       = A_r\rfloor {I}^{-1} 
       = \langle A_r {I}^{-1}\rangle_{n-r}.
\ee 
The OPNS representation by $A_r$ and the dual IPNS representation by $A_r^{\ast}$ are directly related by duality  
\begin{align}
  \forall x\in \R^{p,q}:\,\,\,
  x \rfloor A_r^{\ast} 
  = x \rfloor (A_r\rfloor{I}^{-1})
  = (x \wedge A_r)\rfloor{I}^{-1} 
  = (x \wedge A_r){I}^{-1} ,
\end{align}
which holds again because of \eqref{eq:ABCrel}.
Therefore we have $\forall x \in \R^{p,q}$
\be 
   x \wedge A_r = 0
   \quad \Leftrightarrow \quad x \rfloor  A_r^{\ast} =0 .
\ee

\subsection{Geometric algebra of projective space}

For details of the geometric algebra of projective space, please see: chapter 10.1 of \cite{DL:GAfPh}, chapter 4.2 of \cite{CP:GAwAE}, and more general chapter 2 of \cite{LH:IAaGR}. In the following we can only introduce the most essential notions.  
A projective space (or homogeneous space) consists of equivalence classes of points which form projection rays. It is generated by regarding the elements of $\gvec{a}\in \R^{p,q}$ as equivalence classes $[\gvec{a}]=\{\alpha \gvec{a}: \alpha \in \R \setminus\{0\}\}$, i.e. the line through the origin of $\R^{p,q}$ and $\gvec{a}$, but excluding the origin itself. To generate representatives of equivalence classes $[\gvec{a}]$, the vector $\gvec{a}\in \R^{p,q}$ is embedded  in the affine space of $\R^{p,q}$
\be 
  \gvec{a}\in \R^{p,q} \mapsto a=\gvec{a}+e_0 \in \R^{p+1,q},
\ee 
where the origin of $\R^{p,q}$ is mapped (lifted) to the additional orthonormal vector $e_0: e_0^2=1$. Please note the deliberate use of different fonts for $a \in \R^{p+1,q}$ and $\gvec{a}\in \R^{p,q}$. This embedding is also called homogenization, since any multiple $\alpha a, \alpha \in \R\setminus \{0\}$, reproduces the same $\gvec{a}\in \R^{p,q}$ by
\be 
  \gvec{a} = \frac{\alpha a}{(\alpha a)\rfloor e_0}-e_0
           = \frac{\alpha a}{\alpha}-e_0
           = \gvec{a}+e_0-e_0,
  \label{eq:revembed}
\ee
where $\alpha \in \R\setminus \{0\}$ is called \textit{weight}.
Homogeneous vectors $\gvec{d}\in \R^{p,q}$ that have no $e_0$ component map under $\gvec{d} \mapsto \gvec{d}/({\gvec{d}\rfloor e_0})$ to infinity, they are called points at infinity or direction vectors. 
All multiples of $\alpha a, \alpha \in \R\setminus \{0\}$, are given in the geometric algebra of projective space $Cl_{p+1,q}$ by the OPNS representation
$\{x\in\R^{p+1,q}: x\wedge a=0 \}$. The point $P$ at position $\gvec{p}\in \R^{p,q}$ is thus represented by a homogeneous vector $p\in \R^{p+1,q}$, i.e. the OPNS of $p$ in $Cl_{p+1,q}$. This includes points at infinity. 

\begin{figure}
\centering
\includegraphics[scale=0.4]{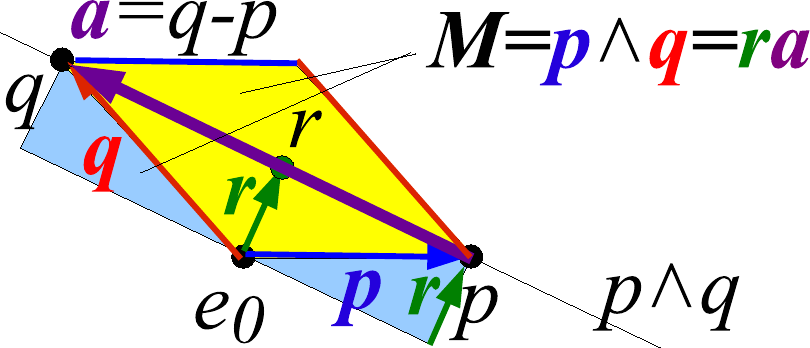}
\caption{The rectangularly reshaped moment bivector of a line is the geometric product of the support vector times the direction vector $\gvec{M}=\gvec{r}\gvec{a}$. \label{fg:fig6}}
\end{figure}

We now investigate lines in $n$-dimensional Euclidean space modeled in the projective geometric algebra $Cl(\R^{n+1})=Cl_{n+1}$. In both Grassmann algebra and in Clifford's geometric algebra the subspace spanned by two linearly independent vectors is given by their outer product. And in projective geometric algebra indeed the outer product of two points spans (in OPNS) the line through these two points, including its offset from the origin $e_0$
\be 
  \label{eq:Lbiv}
  p \wedge q 
  = e_0\wedge (\gvec{q}-\gvec{p}) + \gvec{p}\wedge \gvec{q}
  = e_0\wedge \gvec{a} + \gvec{M},
\ee 
where we used unit weight points (scalar multiples would span the same subspace). 
For any point $x=\alpha p + \beta q, \forall \alpha,\beta \in \R$ on the line has zero outer product with $p \wedge q$
\be 
  x \wedge (p\wedge q) = 0 \,\,\,
  \Leftrightarrow \,\,\,
  x = \alpha p + \beta q, \,\,\, \alpha,\beta \in \R.
\ee 
We recognize $\gvec{a}=\gvec{q}-\gvec{p}\in\R^n$ as the direction vector of the line. Reshaping the Euclidean bivector $\gvec{M}=\gvec{p}\wedge \gvec{q}$ with the Gram-Schmidt process to rectangular shape (see Fig. \ref{fg:fig6})
\be 
  \label{eq:LMoment}
  \gvec{M}
  = \gvec{p}\wedge (\gvec{q}-\gvec{p})
  = \gvec{r} (\gvec{q}-\gvec{p})\,\,\,
  \Leftrightarrow
  \,\,\,
  \gvec{r} = (\gvec{p}\wedge \gvec{q}) (\gvec{q}-\gvec{p})^{-1}
           = \gvec{M}\gvec{a}^{-1},
\ee 
we find it to be the \textit{moment} (bivector) $\gvec{M}=\gvec{r}\gvec{a}$ of the line, i.e. the geometric product of its distance vector $\gvec{r}\in\R^n$ from the origin, times $\gvec{a}$. $\gvec{r}$ is also called (perpendicular) support vector (relative to the origin $e_0$) with projective point representation 
\be 
  r = e_0 + \gvec{r} \in \R^{n+1}.
\ee 
An alternative specification is the point-direction representation (or line-bound vector from an affine perspective) of the line space bivector
\be 
  p\wedge q = p \wedge (q-p) = p \wedge \gvec{a},
\ee 
where $p$ is a point on the line and $\gvec{a}$ the direction vector.

\subsection{Conformal model of Euclidean space}

The conformal model \cite{DFM:GAfCS}
of Euclidean space (in the GA of $\mathbb{R}^{n+1,1}$) 
adds two null-vector dimensions for 
the origin ${e}_0$ and infinity ${e}_{\infty}$, such that
\begin{align}
  \forall \gvec{x} \in \R^n: \quad
  X = \gvec{x} + \frac{1}{2}\gvec{x}^2{e}_{\infty}+{e}_0 \in \R^{n+1,1}, 
\qquad
  {e}_0^2 = {e}_{\infty}^2=X^2=0,
\qquad
  e_0\rfloor \einf = X\rfloor {e}_{\infty} = -1.
\end{align}
The $+{e}_0$ term integrates projective geometry,
and the $+ \frac{1}{2}\gvec{x}^2{e}_{\infty}$ term
ensures $X^2=0$. The vector space $\mathbb{R}^{n}$ is thus expanded by 
a 2D plane described in OPNS by the 2-blade $E=\einf \wedge e_0$, $E^2=1$.
The inner product of two conformal points $X,A$ gives their Euclidean distance
\be 
  X \rfloor A = -\frac{1}{2}(\gvec{x}-\gvec{a})^2.
\ee 
Therefore (in IPNS) a (mid)plane $\mu$ (actually a hyperplane) equidistant from two points $A,B$ is
\begin{align}
  X\rfloor A = X\rfloor B \quad
  \Rightarrow \quad 
  X\rfloor \mu = X\rfloor (A-B)=0.
\end{align}
And we obtain the vector $\mu$ representing the plane as
\begin{align}
  \mu=A-B &\propto \gvec{n}+d\,{e}_{\infty},
\end{align}
where $\gvec{n}$ is a unit normal to the plane and $d$ its signed scalar distance from the origin. 

Reflecting at two parallel planes $\mu,\mu^{\prime}$ with 
distance $\gvec{t}/2$ we get the so-called \textit{translator} 
(\textit{transla}tion opera\textit{tor} by $\gvec{t}\,$)
\begin{equation}
  X^{\prime} 
  = \mu^{\prime}\mu\,X\,\mu\mu^{\prime} 
  = T_{\gvec{t}}^{-1} X T_{\gvec{t}},
  \quad T_{\gvec{t}}=1+\frac{1}{2}\gvec{t}{e}_{\infty}.
\end{equation}
Reflection at two non-parallel planes $\mu,\mu^{\prime}$ yields the rotation around
the $\mu,\mu^{\prime}$-intersection axis (a line for $n=3$) by twice the angle subtended by $\mu,\mu^{\prime}$.

An alternative dual description \cite{EH:EucGObj,HTBY:Carrier} of 2D planes is found in the OPNS representation by wedging three conformal points $A,B,C \in \R^{n+1,1}$ on the plane with infinity
\be 
  \Pi 
  = A\wedge  B \wedge C \wedge \einf
  = \mathbf{D}\gvec{d}\einf -\mathbf{D}E,
  \quad 
  \mathbf{D} = (\gvec{a} - \gvec{b}) \wedge (\gvec{b} - \gvec{c}),
\ee 
where the Euclidean bivector $\mathbf{D}$ gives the direction 2-blade of the plane and $\gvec{d}\in \R^{n}$ the shortest distance vector of the plane from the origin (support vector). Direct computation shows that $\Pi^2 = \mathbf{D}^2 = -|\mathbf{D}|^2$ and we can therefore norm the conformal plane 4-vector by $\Pi \rightarrow \Pi/|\mathbf{D}|$.

Group theoretically the conformal group $C(n)$ is isomorphic to $O(n+1,1)$ and the
Euclidean group $E(n)$ is the subgroup of $O(n+1,1)$ leaving infinity 
${e}_{\infty}$ invariant.
Now general translations and rotations are represented by geometric products
of vectors. For an application of these concepts in $Cl_{4,1}$ to interactive crystal symmetry visualization see \cite{EH:Grassmann200, PH:SGV, HP:ICCA8}.

\section{Crystal planes in geometric algebra of projective space}
\subsection{Barycentric coordinates versus fractional coordinates}

We first consider the \textit{barycentric coordinates} of projective geometry and their relationship to the \textit{fractional coordinates} of crystallography. 

\begin{figure}
\centering
\includegraphics[scale=0.5]{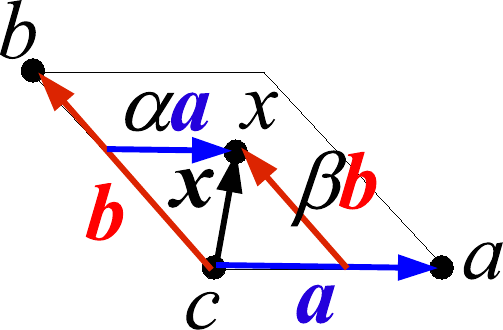}
\caption{Fractional coordinates $\alpha$ and $\beta$ in two dimensions. \label{fg:fig1}}
\end{figure}

\begin{figure}
\centering
\includegraphics[scale=0.5]{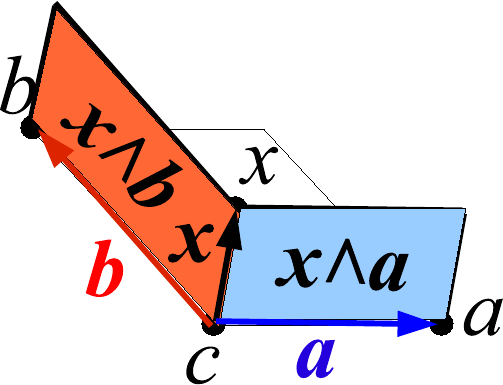}
\includegraphics[scale=0.5]{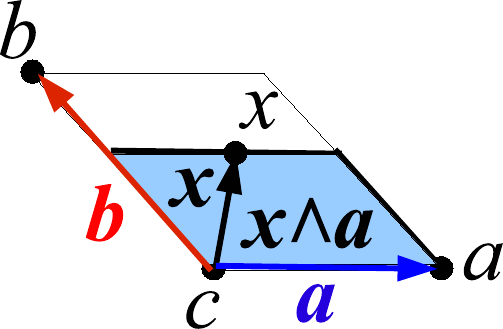}
\includegraphics[scale=0.5]{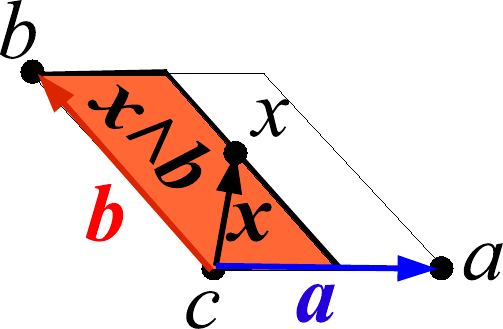}
\caption{Left: Oriented area bivectors $\gvec{x} \wedge \gvec{a}$ and  $\gvec{x} \wedge \gvec{b}$. 
Center: Reshaped bivector $\gvec{x} \wedge \gvec{a}$. 
Right: Reshaped bivector $\gvec{x} \wedge \gvec{b}$.
\label{fg:fig2}}
\end{figure}

\begin{figure}
\centering
\includegraphics[scale=0.5]{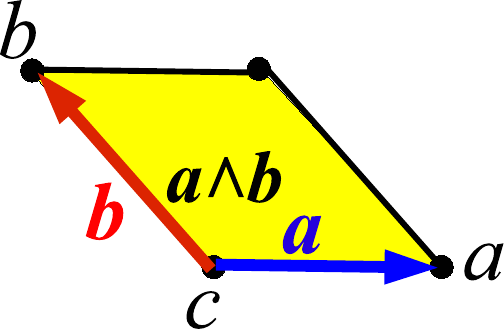}
\caption{Two-dimensional crystal cell spanned by vectors $\gvec{a}, \,\gvec{b}$. \label{fg:fig3}}
\end{figure}

In an offset 2D plane a point $x\in\R^{n+1}$ can be represented as a linear combination of three points $a, b, c \in\R^{n+1}$ in general location. (We can set $n=3$, but our results are valid for general $n$.) If the three points are of unit weight, the linear combination becomes an \textit{affine combination}
\be 
  x = \alpha a + \beta b + \gamma c, \quad \alpha + \beta + \gamma = 1, 
  \quad \alpha, \beta, \gamma \in \R. 
\ee 
We compute the coefficients by subtracting $c$ on both sides and wedging with $\gvec{a} = a-c \in\R^n$ and $\gvec{b} = b-c\in\R^n$, respectively (see Fig. \ref{fg:fig1})
\be 
  \gvec{x} = x-c = \alpha \gvec{a} + \beta \gvec{b} + (\alpha+\beta+\gamma-1)c
  = \alpha \gvec{a} + \beta \gvec{b} \in\R^n.
\ee 
We obtain $\beta$ as a ratio of oriented areas (see Fig. \ref{fg:fig2}) by
\be 
  \gvec{x} \wedge \gvec{a} = \beta \gvec{b}\wedge \gvec{a}
  \Rightarrow 
  \beta =  \frac{\gvec{x} \wedge \gvec{a}}{\gvec{b}\wedge \gvec{a}}
  = -\frac{\gvec{x} \wedge \gvec{a}}{\gvec{a}\wedge \gvec{b}},
  \,\,\,\text{ and similarly }\,\,\,
  \alpha = \frac{\gvec{x} \wedge \gvec{b}}{\gvec{a}\wedge \gvec{b}},
\ee 
and therefore
\be 
  \gamma = 1 - \alpha - \beta 
  = 1- \frac{\gvec{x} \wedge \gvec{b}}{\gvec{a}\wedge \gvec{b}}
     + \frac{\gvec{x} \wedge \gvec{a}}{\gvec{a}\wedge \gvec{b}}
  =  1- \frac{\gvec{x} \wedge (\gvec{b}-\gvec{a})}{\gvec{a}\wedge \gvec{b}}.
\ee 
In the denominator of $\alpha$ and $\beta$ we have the oriented volume (area) $\gvec{a}\wedge \gvec{b} \in \bigwedge^2\R^n$ of a cell (parallelogram spanned by $\gvec{a}$ and $\gvec{b}$), see Fig. \ref{fg:fig3}. Comparing Figs.  \ref{fg:fig2} (Right) and  \ref{fg:fig3}, the bivector area ratio $(\gvec{x} \wedge \gvec{b})/(\gvec{a} \wedge \gvec{b})$ obviously gives $\alpha$ of Fig. \ref{fg:fig1}. Similarly comparing Figs.  \ref{fg:fig2} (Center) and  \ref{fg:fig3} we see that the bivector area ratio $-(\gvec{x} \wedge \gvec{a})/(\gvec{a} \wedge \gvec{b})$ gives $\beta$ of Fig. \ref{fg:fig1}. The minus sign in the denominator is due to the opposite orientations of $\gvec{x} \wedge \gvec{a}$ (clockwise) and $\gvec{a} \wedge \gvec{b}$ (anti-clockwise).
In the crystallography of 2D crystals, point $c$ is often called the origin of the cell, and the coordinate values $\alpha$ and $\beta$ are called \textit{fractional coordinates}. 

The \textit{barycentric coordinates} $\alpha, \beta, \gamma$ can be used to interpolate a scalar property $S: \R^n\rightarrow \R$ given at the vertexes $a, b, c$ to a value at $x$: $S_x = \alpha S_a + \beta S_b + \gamma S_c$.

\subsection{Reciprocal vectors of crystallography}

\begin{figure}
\centering
\includegraphics[scale=0.5]{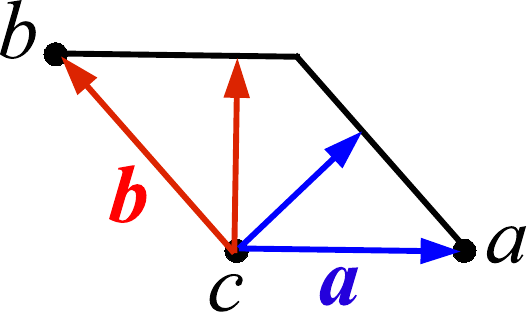}
\caption{Geometrical interpretation of inverse of reciprocal vectors. \label{fg:fig4}}
\end{figure}

We further observe, that the vectors 
\be 
  \gvec{a}' = \gvec{b}/(\gvec{a}\wedge \gvec{b})
            = \gvec{b}\rfloor(\gvec{a}\wedge \gvec{b})^{-1}, \quad
  \gvec{b}' = -\gvec{a}/(\gvec{a}\wedge \gvec{b})
            = -\gvec{a}\rfloor(\gvec{a}\wedge \gvec{b})^{-1}
\ee 
are also called \textit{reciprocal vectors} in crystallography. The reciprocal vectors have the property that
\be 
  \gvec{a}\rfloor \gvec{a}' 
  = \gvec{a}\rfloor (\gvec{b}(\gvec{a}\wedge \gvec{b})^{-1})
  = (\gvec{a}\wedge \gvec{b})(\gvec{a}\wedge \gvec{b})^{-1} = 1,
\ee 
by applying \eqref{eq:ABCrel} for the second equality. 
Similarly
\be 
  \gvec{b}\rfloor \gvec{b}' =1, 
  \quad
  \gvec{a}\rfloor \gvec{b}' 
  = \gvec{a}\rfloor [-\gvec{a}(\gvec{a}\wedge \gvec{b})^{-1}]
  = -(\gvec{a}\wedge \gvec{a})(\gvec{a}\wedge \gvec{b})^{-1} = 0,
  \quad
  \gvec{b}\rfloor \gvec{a}' =0.
\ee 
Since, by applying again \eqref{eq:ABCrel} we can conversely rewrite the coefficient equations as
\be 
  \alpha = \frac{\gvec{x} \wedge \gvec{b}}{\gvec{a}\wedge \gvec{b}}
         = (\gvec{x} \wedge \gvec{b})\rfloor (\gvec{a}\wedge \gvec{b})^{-1}
         = \gvec{x} \rfloor [\gvec{b}/({\gvec{a}\wedge \gvec{b}})]
         = \gvec{x} \rfloor \gvec{a}',
  \,\,\,\text{ and similarly }\,\,\,
  \beta = \gvec{x} \rfloor \gvec{b}',
\ee 
where we observe the familiar role of reciprocal vectors in crystallography.
Note that geometrically the inverse of $\gvec{a}'$ is the rejection of $\gvec{a}$ from $\gvec{b}$
\be 
  \gvec{a}'^{-1} = (\gvec{a}\wedge \gvec{b})\gvec{b}^{-1} = P^{\perp}_{\sgvec{b}}(\gvec{a}),
\ee 
which can be interpreted as the perpendicular distance vector of $a$ from the line $c\wedge b$.
And likewise
\be 
  \gvec{b}'^{-1} = (\gvec{b}\wedge \gvec{a})\gvec{a}^{-1} = P^{\perp}_{\sgvec{a}}(\gvec{b}),
\ee 
the perpendicular distance vector of $b$ from the line $c\wedge a$, see Fig. \ref{fg:fig4}.

\begin{figure}
\centering
\includegraphics[scale=0.5]{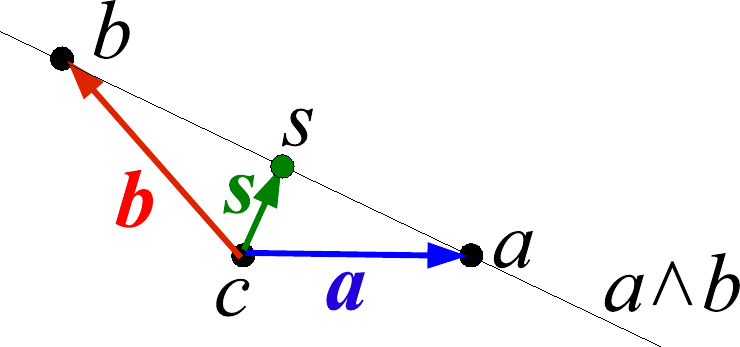}
\caption{Sum of reciprocal vectors $\gvec{a}' + \gvec{b}'$ yields support vector of line $a\wedge b$. \label{fg:fig5}}
\end{figure}

If we add the two reciprocal vectors we obtain 
\be 
  \gvec{a}' + \gvec{b}' 
  = \gvec{b}/(\gvec{a}\wedge \gvec{b})
   -\gvec{a}/(\gvec{a}\wedge \gvec{b})
  = (\gvec{b}-\gvec{a})/(\gvec{a} \wedge \gvec{b})
  = \gvec{s}^{-1},
  \label{eq:2DaddRcVec}
\ee 
i.e. the inverse of the support vector $\gvec{s}\in\R^n$ (relative to point $c$) of the line $a\wedge b = (c+\gvec{s})(\gvec{b}-\gvec{a})$ with direction vector $\gvec{b}-\gvec{a}$, see Fig. \ref{fg:fig5}. 

\section{Generalization to 3D and $k$D $(k\leq n)$ cases}

\subsection{3D case}

The above example of an offset 2D plane spanned by 3 points can easily be generalized to barycentric coordinates in a 3D subspace spanned by 4 points $a,b,c,d \in \R^{n+1}$. If these points are unit-weight points, then an affine combination represents every other point 
$x=\alpha a + \beta b + \gamma c + \delta d$. With analogous definitions $\gvec{x} = x-d \in \R^{n}$, etc., to the plane case the first three coefficients $\alpha, \beta, \gamma$ will be the usual \textit{fractional coordinates} of crystallography for a 3D crystal cell with origin $d$. 

The result is that in the 3D case the first three barycentric coordinates of $x$, corresponding exactly to the fractional coordinates of crystallography, are given by
\be 
  \alpha = \frac{\gvec{x} \wedge \gvec{b} \wedge \gvec{c}}{\gvec{a}\wedge \gvec{b}\wedge \gvec{c}}, \quad
  \beta = \frac{\gvec{a} \wedge \gvec{x} \wedge \gvec{c}}{\gvec{a}\wedge \gvec{b}\wedge \gvec{c}}, \quad 
  \gamma = \frac{\gvec{a} \wedge \gvec{b} \wedge \gvec{x}}{\gvec{a}\wedge \gvec{b}\wedge \gvec{c}},
\ee 
where $\gvec{a} = a - d, \gvec{b}=b-d, \gvec{c}= c-d$.

The \textit{reciprocal vectors} of a 3D cell with cell vectors $\gvec{a}, \gvec{b}, \gvec{c}$ are 
\be
  \gvec{a}' = \frac{\gvec{b} \wedge \gvec{c}}{\gvec{a}\wedge \gvec{b}\wedge \gvec{c}}, \quad 
  \gvec{b}' = \frac{\gvec{c} \wedge \gvec{a}}{\gvec{a}\wedge \gvec{b}\wedge \gvec{c}}, \quad
  \gvec{c}' = \frac{\gvec{a} \wedge \gvec{b}}{\gvec{a}\wedge \gvec{b}\wedge \gvec{c}},
\ee 
where the denominator $\gvec{a}\wedge \gvec{b}\wedge \gvec{c}$ is the oriented 3-volume of the cell, and the numerator is obtained by removing the vector, whose reciprocal vector is to be defined. 
The inverse reciprocal vectors are heights of corresponding points over a side face of the parallelepiped cell
\be 
  \gvec{a}'^{-1}=P_{\sgvec{b}\wedge\sgvec{c}}^{\perp}(\gvec{a}), 
  \quad
  \gvec{b}'^{-1}=P_{\sgvec{c}\wedge\sgvec{a}}^{\perp}(\gvec{b}),
  \quad
  \gvec{c}'^{-1}=P_{\sgvec{a}\wedge\sgvec{b}}^{\perp}(\gvec{c}),
\ee 
i.e. the height vectors of $a$ over side plane $b \wedge c \wedge d$, 
of $b$ over $c \wedge a \wedge d$, and $c$ over 
$a \wedge b \wedge d$, respectively.
The reciprocal vectors fulfill 
\be 
  \gvec{a}\rfloor \gvec{a}' 
  = \gvec{b}\rfloor \gvec{b}' 
  = \gvec{c}\rfloor \gvec{c}' =1, \quad
  \text{ while all other products vanish } \quad
  \gvec{b}\rfloor \gvec{a}' = \gvec{c}\rfloor \gvec{a}' = 0, \,\ldots
\ee 

We further observe that adding the three reciprocal vectors gives the inverse of the support vector (relative to point $d$) of the plane 
$a\wedge b\wedge c = s\gvec{A}, \,\,s = (d+\gvec{s}) \in \R^{n+1}$, 
\be 
  \gvec{s}^{-1} 
  = \gvec{a}'+\gvec{b}'+\gvec{c}'
  = \frac{\gvec{b} \wedge \gvec{c}+\gvec{c} \wedge \gvec{a}+\gvec{a} \wedge \gvec{b}}{\gvec{a}\wedge \gvec{b}\wedge \gvec{c}}
  = \gvec{A}/\gvec{M},
\ee  
with direction bivector $\gvec{A}=\gvec{b} \wedge \gvec{c}+\gvec{c} \wedge \gvec{a}+\gvec{a} \wedge \gvec{b}$ and moment trivector $\gvec{M}=\gvec{a}\wedge \gvec{b}\wedge \gvec{c}$. The expression $s\gvec{A}$ is the support (point) $\times$ direction form of the offset plane $a \wedge b \wedge c$.

\subsection{Generalization to offset $k$D $(k\leq n)$ subspaces }

The concept of affine combination of $k+1$ (unit-weight) points $a_1, \ldots, a_k, a \in \R^{n+1}$ to determine any other point $x$ in the offset $k$D subspace thus spanned leads to the corresponding barycentric and fractional coordinates, as well as the reciprocal vectors ($\check{\gvec{a}}_l$ means to omit $\gvec{a}_l=a_l-a$)
\be 
  \gvec{a}{'}_l= (-1)^{l-1}(\gvec{a}_1\wedge \ldots \check{\gvec{a}}_l\ldots\wedge \gvec{a}_k)/(\gvec{a}_1\wedge \ldots \wedge \gvec{a}_k),
\ee
of a $k$D parallelepiped cell of crystallography. The inverse of each reciprocal vector
\be
 \gvec{a}{'}_l^{-1} = P_{(-1)^{l-1}\sgvec{a}_1\wedge \ldots \check{\sgvec{a}}_l\ldots\wedge \sgvec{a}_k}^{\perp}(\gvec{a}_l)
\ee
is again the rejection of $\gvec{a}_l$ from the $(k-1)$D side face 
$\gvec{a}_1\wedge \ldots \check{\gvec{a}}_l\ldots\wedge \gvec{a}_k$, i.e. the
height vector of point $a_l$ over the side face $(k-1)$D plane ${a}_1\wedge \ldots \check{{a}}_l\ldots\wedge {a}_k\wedge a$.
Vectors and reciprocal vectors are related by
\be 
  \gvec{a}_m\rfloor \gvec{a}{'}_l = \delta_{m,l}, \quad 1\leq m,l \leq k.
\ee
Any vector $x$ in the offset $k$D subspace spanned by $a_1, \ldots a_k, a$ can be represented (by affine combination) as
\begin{align} 
  \hspace{-2mm}x = \hspace{-1mm}\sum_{l=1}^k (\gvec{x}\rfloor \gvec{a}{'}_l) a_l 
      + (1-\hspace{-1mm}\sum_{l=1}^k \gvec{x}\rfloor \gvec{a}{'}_l) a
    = a+\hspace{-1mm}\sum_{l=1}^k (\gvec{x}\rfloor \gvec{a}{'}_l) (a_l-a)
  \Leftrightarrow 
  \gvec{x} = x-a = \hspace{-1mm}\sum_{l=1}^k (\gvec{x}\rfloor \gvec{a}{'}_l) \gvec{a}_l .
  \label{eq:PoffsSubspParEq}
\end{align}
The first $k$ barycentric coordinates $\gvec{x}\rfloor \gvec{a}{'}_l, 1\leq l \leq k$ 
correspond thus to the fractional coordinates of a $k$D crystallographic cell, projectively embedded in $\R^{n+1}$. Equation \eqref{eq:PoffsSubspParEq} represents a \textit{parametric equation} of a $k$D offset subspace in projective geometric algebra, to be compared with the $(k+1)$-blade representation $a_1 \wedge \ldots \wedge a_k \wedge a$.

\section{Crystal hyperplanes}
\subsection{$(k-1)$D subspaces and hyperplanes}

The duality operation of $Cl(\R^{n+1})$ is defined as multiplication by the inverse pseudoscalar $(\gvec{I}a)^{-1}$ of $Cl(\R^{n+1})$. Here $\gvec{I}$ is the pseudoscalar of the Euclidean space subalgebra $Cl(\R^{n})\subset Cl(\R^{n+1})$ and the $a\in\R^{n+1}$. For simplicity we assume $a$ to be of unit weight.

The support vector of the $(k-1)$D offset subspace $\Pi_{k-1}=a_1\wedge \ldots \wedge a_k$ relative to point $a$ is given by
\be 
  \gvec{d}^{-1} = \sum_{l=1}^k \gvec{a}{'}_l.
\ee 

If the offset subspace is a hyperplane, i.e. $k=n$, then its \textit{dual} representation (as inner product null space [IPNS]) is given by the point 
\be 
  \pi =\Pi_{n-1}(\gvec{I}a)^{-1}= a-\gvec{d}^{-1}=a-\sum_{l=1}^n \gvec{a}{'}_l \in \R^{n+1}.
\ee

\subsection{Space lattices and Miller indexes}

Let the points $a_1, \ldots a_n, a$ represent an $n$D crystal cell, with $a$ as the origin of the cell. 
Applying repeatedly the translations $T_{\sgvec{a}_l}[\,], 1\leq l \leq n$ forms a space lattice.
Any hyperplane of this space lattice 
 can be dually represented by points $\pi$ (setting $a=e_0$ for simplicity)
\be 
  \pi_{(h_1, \ldots, h_n)} 
  = e_0-\gvec{d}^{-1}_{(h_1, \ldots, h_n)} 
  = e_0-\sum_{l=1}^n h_l \gvec{a}{'}_l,
  \quad 
  \forall\, h_l\in \Z, 1\leq l \leq n.
\ee 
For $n=3$ the (relatively prime) integer coefficients are usually called \textit{Miller indexes}
\be 
  \pi_{(h k l)} 
  = e_0-\gvec{d}^{-1}_{(h k l)} 
  = e_0- (h \gvec{a}{'}_1 + k \gvec{a}{'}_2+l \gvec{a}{'}_3),
  \quad 
  \forall\, h,k,l \in \Z.
\ee 

\begin{example}[Plane with $(hkl)=(1,3,2)$]
Assume a plane with Miller indexes $(hkl)=(1,3,2)$. Then we can immediately write down the dual point form of the plane
\be 
  \pi_{(1, 3, 2)} 
  = e_0-\gvec{d}^{-1}_{(1, 3, 2)} 
  = e_0- (1 \gvec{a}{'} + 3 \gvec{b}{'}+2 \gvec{c}{'}).
\ee 
We can compute the distance vector between two neighboring planes as
\be 
  -\gvec{d}^{-1}_{(1, 3, 2)} 
  = e_0\rfloor(e_0\wedge \pi_{(1, 3, 2)})
  \Rightarrow 
  \gvec{d}_{(1, 3, 2)} = -[e_0\rfloor(e_0\wedge \pi_{(1, 3, 2)})]^{-1}
\ee 
\end{example}
The scalar distance between two neighboring planes 
(called $d$-\textit{spacing} in crystallography) is given by the length $|\gvec{d}|$ of the support vector $\gvec{d}$
\be 
  d_{hkl} 
  = |\gvec{d}_{(hkl)}| = |e_0\rfloor(e_0\wedge \pi_{(hkl)})|^{-1}. 
\ee

\section{Crystal planes in conformal geometric algebra}
\subsection{Crystal planes in conformal model of Euclidean space
}

We now work in the conformal model of Euclidean space in $Cl_{3+1,1}=Cl_{4,1}$.
A 2D crystal plane $(hkl)$ through three conformal points
\begin{align} 
  A_h = \eo + \gvec{a}/h + \hf \frac{\gvec{a}^2}{h^2} \einf, \,\,\,
  B_k = \eo + \gvec{b}/k + \hf \frac{\gvec{b}^2}{k^2} \einf, \,\,\, 
  C_l = \eo + \gvec{c}/l + \hf \frac{\gvec{c}^2}{l^2} \einf, 
  \label{eq:canpoints}
\end{align} 
is given (in OPNS) by the conformal 4-blade
\be 
  \Pi_{(hkl)} = A_h\wedge  B_k \wedge C_l \wedge \einf.
  \label{eq:hklplane}
\ee 
The reciprocal (dual) conformal vector representation of the $(hkl)$ plane (IPNS) is
\be 
  \pi_{(hkl)}= \Pi_{(hkl)}^{\ast} = \gvec{d}_{hkl}^{-1} + \einf ,
\ee 
normed such that the scalar factor of $\einf$ is $+1$, which is achieved by 
$\pi_{(hkl)} \rightarrow \pi_{(hkl)}/(-\pi_{(hkl)}\rfloor e_0)$. 

NB: Equation \eqref{eq:canpoints} is the canonical choice of axis intersection points, yet perfectly valid expressions for the plane 4-vector (OPNS) or the reciprocal vector (IPNS) of the plane are obtained using any set of three conformal points $A,B,C$ in general position on the plane.

\subsection{Bragg reflections and interfacial angles}

The meet of any point $P=\eo + \gvec{p} + \hf \gvec{p}^2 \einf$ with the plane $\Pi_{(hkl)}$ gives the distance of $P$ from $\Pi_{(hkl)}$ in units of the $d$-spacing of $\Pi_{(hkl)}$
\be 
  P \vee \Pi_{(hkl)} 
  = \Pi_{(hkl)}^{\ast} \lfloor P
  = (\gvec{d}_{hkl}^{-1} + \einf)\lfloor (\eo + \gvec{p} + \hf \gvec{p}^2 \einf)
  = \gvec{d}_{hkl}^{-1}\ast \gvec{p} -1.
\ee 
The meet $P \vee \Pi_{(hkl)} $ allows therefore to directly compute the \textit{phase angle} in the \textit{structure factor} $F_{hkl}$ of an atom at point $P$ in a crystal cell as
\be 
  \phi = 2\pi P \vee \Pi_{(hkl)}.
\ee  

In standard crystallography the scattering power of a stack of netplanes $(hkl)$ of a crystal structure is expressed by the unit cell structure factor 
\be 
  F_{hkl} 
  = \sum_{m=1}^N f_m e^{j\phi_m} 
  =  \sum_{m=1}^N f_m e^{j2\pi P_m \vee \Pi_{(hkl)}},
  \quad
  j \in \C, \,\,\,j^2 = -1,
  \label{eq:structf}
\ee 
where the $f_m$ are the atomic scattering factors of the $N$ individual atoms located at points $P_m$ in the unit cell of the crystal structure. \cite{HW:SymCryst} Since the unit pseudoscalar $I_5 = i_3E$, $i_3 = e_1 e_2 e_3,$ of $Cl_{4,1}$ squares itself to $I_5^2 = i_3^2 E^2 = i_3^2 = -1$, we can replace the complex unit $j$ in \eqref{eq:structf} by $I_5$
\be 
  F_{hkl} 
  =  \sum_{m=1}^N f_m e^{2\pi I_5 P_m \vee \Pi_{(hkl)}}.
  \label{eq:structfI5}
\ee 
The phase angles $2\pi I_5 P_m \vee \Pi_{(hkl)}$ of \eqref{eq:structfI5} can be rewritten in $Cl_{4,1}$ as (we drop the indexes $(hkl)$ and $m$ for simplicity)
\be 
  I_5 P \vee \Pi 
  = I_5 (\Pi^{\ast}\lfloor P)
  = I_5 \left( (\Pi \rfloor I_5^{-1})\lfloor P \right)
  = I_5 \left( P\rfloor (\Pi \rfloor I_5^{-1}) \right)
  = I_5 \left( (P \wedge \Pi) I_5^{-1}) \right)
  = P \wedge \Pi,
\ee 
where we used 
$(\Pi \rfloor I_5^{-1}) \lfloor P = P \rfloor (\Pi \rfloor I_5^{-1})$, because both $(\Pi \rfloor I_5^{-1})$ and $P$ are vectors in $\R^{4,1}$, and \eqref{eq:ABCrel}. This leads to a final fully geometric expression for the structure factor as
\be 
  F_{hkl} 
  =  \sum_{m=1}^N f_m e^{2\pi P_m \wedge \Pi_{(hkl)}},
  \label{eq:structfgeom}
\ee
which offers a very direct method of structure factor computation solely based on the three points $A_h, B_k, C_l$ that define the crystal plane \eqref{eq:hklplane} under consideration. Equations \eqref{eq:structf}, \eqref{eq:structfI5}  or \eqref{eq:structfgeom} yield the \textit{integral Bragg reflection conditions} for centered crystal cells. \cite{HW:SymCryst} 

Selecting a general  point $P$ with position vector
$\gvec{p}=x\gvec{e}_1+y\gvec{e}_2+z\gvec{e}_3$
and subjecting it to a space group symmetry operation given by a conformal versor $V: P \mapsto P' = Ad_V(P) = \widehat{V}^{-1} P V$, see \cite{LH:IAaGR},
allows to compute the phase angle of the symmetrical position $P'$
\be 
  \phi' = 2\pi P' \vee \Pi_{(hkl)},
  \quad \text{or as} \quad 
  I_5 \phi' = 2\pi P'\wedge \Pi_{(hkl)}.
\ee  
Using the versor expressions for glide reflections \cite{EH:Grassmann200, PH:SGV, HP:ICCA8} and inserting into the structure factors $F_{hkl}$ leads to \textit{zonal Bragg reflection conditions} for space groups with glide reflections, using the versor expressions for screw rotations leads to the \textit{serial Bragg reflection conditions} for space groups with screw rotation symmetry. \cite{HW:SymCryst} 

The \textit{interfacial angle} $\theta$ between two crystal planes $(h_1k_1l_1)$ and $(h_2k_2l_2)$ given by their reciprocal representation vectors
\begin{align}
  \pi_1 = \pi_{h_1k_1l_1}= \gvec{d}_{h_1k_1l_1}^{-1} + \einf = \gvec{d}_1^{-1}+ \einf, \,\,\,
  \pi_2 = \pi_{h_2k_2l_2}= \gvec{d}_{h_2k_2l_2}^{-1} + \einf = \gvec{d}_2^{-1}+ \einf,
\end{align}
 is defined as
\be 
  \cos \theta 
  = \frac{\Pi_1 \ast  \Pi_2}{|\Pi_1| |\Pi_2|}
  = \frac{\pi_1 \ast \pi_2}{|\pi_1| |\pi_2|}
  = \frac{(\gvec{d}_{1}^{-1}+\einf)\ast(\gvec{d}_{2}^{-1}
    +\einf)}{|\gvec{d}_{1}|^{-1} |\gvec{d}_{2}|^{-1}}
  = \frac{\gvec{d}_{1}\ast\gvec{d}_{2}}{|\gvec{d}_{1}| |\gvec{d}_{2}|},
\ee 
since $\gvec{d}_{j}^{-1}= \gvec{d}_{j}/|\gvec{d}_{j}|^2$, $\gvec{d}_{j}\ast \einf=0$, $j=1,2$, and because the factors $|\gvec{d}_{j}|^2$, $j=1,2$, in the numerator and denominator cancel out.

By instead using unit norm reciprocal representation vectors, we can use the counter $m \in \Z$ in
\be 
  \pi_{(hkl)} = \gvec{n}_{(hkl)}+m \,d_{(hkl)}\,\einf, 
  \quad \pi_{(hkl)}^2 = \gvec{n}_{(hkl)}^2=1,
\ee 
to faithfully represent every single plane in the family of crystal planes given by the reciprocal vector $\gvec{d}_{(hkl)}^{-1} = \gvec{n}_{(hkl)}/d_{(hkl)} \in \R^n$. For $m=0$ the plane $\pi_{(hkl)} = \gvec{n}_{(hkl)}$ includes the origin.

\section{Conclusion}

We have explained how 2D, 3D and $k$D cells of crystallography can be described in the geometric algebra $Cl(\R^{n+1})$ of projective space $\R^{n+1}$. This lead to the identification of barycentric coordinates with fractional coordinates of crystallography and to the introduction of reciprocal vectors with geometric interpretation. The role of reciprocal vectors in describing offset lines, offset $(k-1)$D subspaces and offset hyperplanes in projective geometric algebra $Cl(\R^{n+1})$ was studied, including crystal planes. Finally the algebraic framework was expanded to the conformal model of Euclidean geometry in $Cl(\R^{n+1,1})$, which elegantly allows to faithfully represent families of crystal planes, and to directly compute interfacial angles, phase angles of atomic structure factors, as well as integral, zonal and serial conditions for Bragg reflections. It may be of interest in the future to extend this treatment to the theory of composite lattices in crystallography \cite{RMRK:CSL}.

\section*{Acknowledgments}

I do thank God:
\begin{quote}
I therefore believe the truths revealed in the Bible, not because they are written in the Bible, but because I have experienced in my own conscience their power of blessing, their eternal, divine truth. 
(\textit{H. Grassmann 1809}--1877)~\cite{HG:AbvGlauben}
\end{quote}
I further thank
 my wife, my children, my parents,
 C. Perwass, D. Hestenes, M. Aroyo, M. Nespolo, B. Souvignier,
 L. Dorst, D. Fontijne, and J. Lasenby. I finally thank the reviewer for his helpful comments.


\end{document}